\documentclass[12pt,preprint]{aastex}

\newcommand{\be}{\begin{equation}}
\newcommand{\ee}{\end{equation}}

\begin{document}

\title{High-Energy Activity in the Unusually Soft TeV Source \\ 
HESS J1804-216 toward the Galactic Center} 

\medskip
\author{Marco Fatuzzo$^1$, Fulvio Melia$^2$, and Roland M. Crocker$^{3}$}
\bigskip 
\affil{$^1$Physics Department, Xavier University, Cincinnati, OH 45207} 
\affil{$^2$Physics Department and Steward Observatory, 
The University of Arizona, AZ 85721}
\affil{$^3$ School of Chemistry and Physics, The University of Adelaide,
South Australia 5005}

\begin{abstract} 
In recent years, apparent anisotropies in the ~EeV cosmic ray (CR) flux
arriving at Earth from the general direction of the galactic center have
been reported from the analysis of AGASA and SUGAR data. The more recently 
commissioned Auger Observatory has not confirmed these results. HESS has now 
detected an unusually soft TeV source roughly coincident with the location of the 
previously claimed CR anisotropy. In this paper, we develop a model for the 
TeV emission from this object, consistent with observations at other wavelengths, 
and examine the circumstances under which it might have contributed to the 
$\sim$ EeV cosmic ray spectrum. We find that the supernova remnant 
G8.7-0.1 can plausibly account for all the known radiative characteristics
of HESS J1804-216, but that it can accelerate cosmic rays only up to an
energy $\sim 10^5$ GeV. On the other hand, the pulsar (PSR J1803-2137) 
embedded within this remnant can in principle inject EeV protons into
the surrounding medium, but it cannot account for the broadband
spectrum of HESS J1804-216. We therefore conclude that although G8.7-0.1
is probably the source of TeV photons originating from this direction, there
is no compelling theoretical motivation for expecting a cosmic ray anisotropy 
at this location. However, if G8.7-0.1 is indeed correctly identified
with HESS J1804-216, it should also produce a $\sim$ GeV flux detectable in
a one-year all sky survey by GLAST.
\end{abstract}


\section{Introduction}  
The High Energy Stereoscopic System (HESS) array provides sensitivity to gamma 
rays with energy $>100$ GeV at a level
below $1\%$ of the flux from the Crab Nebula, with an angular resolution for
individual photons better than $0.1^\circ$. Thus, the position of even relatively 
faint sources may be determined with an error of only $30$ sec. A scan of the
inner $60^\circ$ of the galactic plane has identified fourteen discrete TeV-emitting 
sources, half of which have plausible identifications at other wavelengths (Aharonian 
et al. 2005). HESS J1804-216 is the source at galactic coordinates $l=8.40^\circ$ and 
$b=-0.033^\circ$, with a TeV-size of $\approx 22$ arcmin. Its estimated flux above 
200 GeV is $5.3\times 10^{-11}$ cm$^{-2}$ s$^{-1}$, with a statistical error of 
$10$ to $35\%$. Its spectral (power-law) index is $-2.72\pm0.06$, the steepest of
all the TeV sources in this survey (Aharonian et al. 2006).

This source coincides with the southwestern rim of the shell-type SNR G8.7-0.1
(W30), whose radio-emitting radius has been set at $\approx 26$ arcmin (see,
e.g., Handa et al. 1988). Its radio flux at $1$ GHz may be derived from the Green 
(2004) catalog, assuming a radio spectral index $\alpha\sim-0.65$, and integrating 
from $10^7$ to $10^{11}$ Hz (Helfand et al. 2005). Such an estimate yields a
radio flux for G8.7-0.1 of $1.1\times 10^{-11}$ ergs cm$^{-2}$ s$^{-1}$.

From CO observations, it is known that the W30 complex is surrounded by molecular 
gas (Blitz, Fich, \& Stark 1982), in which new stars are forming. By associating 
G8.7-0.1 with coincident H II regions whose distances are known, Kassim \& Weiler 
(1990) have estimated a distance to the SNR of $6\pm 1$ kpc which, combined with 
its angular size of $\sim 50$ arcmin, implies a physical size of $\sim 80$ pc. 
Thus, if HESS J1804-216 is indeed associated with this SNR, its location is not 
quite at the galactic center (i.e., at a distance of $\approx 8.5$ kpc). 

G8.7-0.1 may also be linked with the (relatively) young pulsar PSR J1803-2137,
an association suggested by their coincidence on the sky, and by the observed 
dispersion measure, which points to a distance of $\sim 5.3$ kpc (Clifton \& 
Lyne 1986). Both sources were observed with the Position Sensitive Proportional
Counter at the focus of ROSAT (Finley \& \"Ogelman 1994), and both were detected
in the soft X-ray energy band $0.1-2.4$ keV. For G8.7-0.1, the unabsorbed flux
in this range of wavelengths is estimated to be $\sim (1-3)\times 10^{-10}$
ergs cm$^{-2}$ s$^{-1}$, corresponding to a luminosity $(0.4-1.3)\times
10^{36}$ ergs s$^{-1}$ (for a distance of $6$ kpc). On the other hand, there is 
no known detection of either source at $\sim$ GeV energies, implying an upper
limit to their EGRET (i.e., $100$ MeV to $\sim 30$ GeV) flux of $\sim 4\times
10^{-8}$ cm$^{-2}$ s$^{-1}$ (Hartman et al. 1999). 

Several of these new TeV sources have raised important questions concerning their 
origin. HESS J1804-216, in particular, is characterized by an unusually steep
gamma ray spectrum, yet it was not detected by EGRET in the GeV range, as one
might naively expect based on an extrapolation of the HESS data to lower energies.
Clearly, the spectrum cannot be a simple power law, and must display some interesting
physics between GeV and TeV energies, possibly shedding some light on the connection
between particle acceleration, diffusion, and emissivity. Addressing these questions
is one of the principal goals of this paper.

But interest in the high-energy activity in this region of the Galaxy extends 
beyond just these issues (see, e.g., Melia and Falcke 2001). The analysis, 
in recent years, of data from two different 
cosmic ray detectors has suggested the presence of an anisotropic overabundance of 
cosmic rays coming from the general direction of the galactic center at energies 
around an EeV ($10^{18}$ eV).  Statistically, the most robust determination for an 
anisotropy has been made by the Akeno Giant Shower Array (AGASA) Group 
\citep{Hayashida1999}, which found a strong---4\% amplitude---anisotropy in the 
energy range $10^{17.9} - 10^{18.3}$ eV.  Two-dimensional analysis of the data showed 
that this anisotropy could be interpreted as an excess of air showers from two regions 
each of $\sim 20^\circ$ extent, one of 4$\sigma$ significance near the galactic center and 
another of 3$\sigma$ in Cygnus. Interestingly, AGASA also saw a CR deficit towards 
the Galactic anti-center.

Prompted by this result, Bellido et al. (2001) re-analyzed the data collected by 
the SUGAR cosmic ray detector, which operated near Sydney from 1968 to 1979. 
They confirmed the existence of an anisotropy, consistent with a point source 
located 7.5$^\circ$ from the galactic center---and 6$^\circ$ degrees from the 
AGASA maximum over an energy range of $10^{17.9} - 10^{18.5}$ eV.

Not surprisingly, this tentative result has generated some intense theoretical
interest because although it has long been speculated that diffusive shock
acceleration of protons and ions at shock fronts associated with supernova
remnants (SNRs) is the mechanism likely responsible for energizing the bulk of
high energy cosmic rays, the observational evidence for this has been elusive. 
In addition, the conditions at almost all known SNRs seem not to promote the 
acceleration of CRs beyond the `knee' feature in the spectrum, at $\simeq 5 
\times 10^{15}$ eV. Thus, the origin of CRs between the knee and the `ankle' 
at few $\times 10^{19}$ eV has been a mystery.

In a series of earlier papers (Crocker et al. 2005a, 2005b; see also Bossa et al. 
2003, Aharonian and Neronov 2005, and the original discussion in Markoff et al.  1997, 
1999), we explored in detail the physics of particle acceleration and radiative emission 
in the central few parsecs of the Galaxy to examine the extent to which detectors on Earth 
would be able to sense neutrinos originating at the galactic center, and to understand how a cosmic 
ray anisotropy might fit into the overall astroparticle scheme associated with this
dynamic region of the Galaxy.

But there are several serious difficulties that a comprehensive and self-consistent
model of the energetic particle activity at the galactic center must overcome (Melia \& Falcke 
2001; Melia 2006). For example, not all of the data can be accommodated within a solitary 
framework, at least not one focusing on the nucleus itself; indeed, the data are inconsistent 
amongst themselves in two important instances: (i) the SUGAR results indicate a point 
source offset by $7.5^\circ$ (toward positive galactic latitude) from the galactic center, in 
(at least partial) disagreement with AGASA, and (ii) the $\sim$ TeV and (EGRET) 
$\gamma$-ray emissivities are not simple extensions of a single spectrum. 

Even at $\sim$ EeV energies, charged particles would not be able to reach Earth
directly without being significantly deflected by the intervening magnetic field. A 
consensus has developed that, if real, this CR anisotropy would be caused by
neutron emission at the galactic center, an idea first mooted by Jones (1990).  
The anisotropy `turn on' at a definite energy of $\sim$ EeV finds a natural 
explanation in the fact that this energy corresponds to a Lorentz factor for 
neutrons large enough that they can reach us from the center of the Milky Way. 
Neutrons below this energy decay in transit and are then diverted by galactic 
magnetic fields.  Above $\sim 10^{18.4}$ eV, the anisotropy ceases due to either 
a very steep galactic center source spectrum or an actual cut-off in the source 
so that the background takes over again at this energy.

Note, moreover, that the galactic center, with declination $\delta = -28.9^\circ$, 
is outside the
field of view of AGASA (which is limited to $\delta > -24^\circ$; Bossa et al. 2003),
but not so for SUGAR. Thus, the discrepancy between the two source positionings
may simply be due to the fact that AGASA is seeing protons produced during in-flight
neutron decay, whereas SUGAR sees the neutron source directly. That the SUGAR 
anisotropy is not coincident with the galactic center, however, still presents a challenge 
to all scenarios that would posit an EeV source right at the nucleus.  Either, all 
such models are incorrect or the SUGAR directional determination is somewhat in error.
If not the galactic center, then the SUGAR anisotropy could be due to another source
displaced from the galactic center by approximately $8^\circ$ toward positive galactic latitude. 

Another serious problem with the data sets is that whereas the galactic center is outside the 
field of view of AGASA, the position of the SUGAR maximum is inside the AGASA 
field of view so that the putative SUGAR source should be seen by AGASA. Of course, 
one way out of this dilemma is that the source may have varied between the SUGAR 
and AGASA observation times; this might occur, for example, if the source of
energetic particles were a pulsar. 

One cannot ignore the fact that the new TeV source, HESS J1804-216, coincides
with the previously cataloged SNR, G8.7-0.1, at about the position where SUGAR 
detected the source of CR anisotropy. Could this be the smoking gun that finally 
provides the observational evidence of a link between EeV particle acceleration and 
a known source?

Unfortunately, such an evaluation is not quite straightforward. The Pierre Auger Observatory 
has now acquired enough data to search for an excess of events near the direction of the 
galactic center in several energy bands around an EeV. With the accumulated statistics---1155 
events, compared with an expected number of $1160.7$ for the energy range $(1.0-2.5)$ 
EeV---already larger than that of any earlier experiment, including AGASA and SUGAR (the 
event number of AGASA in this region is only $1/3$ of this), the Auger 
Observatory does not confirm the previously reported CR anisotropy (Letessier-Selvon 
et~al. 2005).  This raises several pertinent questions that we wish to address in 
this paper. (1) Though unlikely, could it be that Auger is wrong, and that the
TeV characteristics of HESS J1804-216 would support the view that SUGAR (and possibly 
AGASA) are in fact correct? Finding a model of HESS J1804-216 as the source of TeV gamma 
rays, not violating observed flux limits at other wavelengths, and producing $\sim$ EeV 
particles, could lend important support to this viewpoint. (2) On the other hand, if Auger 
is correct, and no CR anisotropy is evident from the galactic center, then what is the 
nature of HESS J1804-216, and can one account for it using more `conventional' means, 
i.e., by identifying it as a member of a known class of object? (3) Finally, how likely 
is it that HESS J1804-216 may have produced variable EeV emission over the past 
$\sim 30$ years?  In the next section, we shall develop viable models of its high-energy 
activity, and then discuss its role in a broader context in \S\ 3. 

\newpage
\section{Two Possible Sources for the TeV Photons}

\centerline{\sl 2.1 The Pulsar PSR J1803-2137}
Young, rapidly spinning pulsars have long been viewed as viable sources of cosmic rays, 
possibly capable of accelerating hadrons to energies greater than $\sim 10^{20}$ eV 
(Blasi et al. 2000; Arons 2003).  We consider here the possibility that relativistic
particles produced by PSR J1803-2137 lead to a pionic-decay photon emissivity,
accounting for the HESS J1804-216 source. We also ascertain the likelihood that 
PSR J1803-2137 is the source of excess cosmic rays reported by SUGAR and AGASA.

Following Blasi et al. (2000) and Arons (2003), we consider the acceleration of 
charged particles across voltage drops in the relativistic winds near the light cylinder
of young, rapidly rotating neutron stars.  These particles do not suffer from 
the radiation losses that limit their polar cap or outer gap counterparts to energies 
$\sim 10^{16}$ eV (see, e.g., Arons 2003); as such, they can reach a maximum energy of
\be
E_{\rm max} (\Omega_0) \approx \eta\, Ze\, \Phi_{\rm wind} = \eta\, Ze {\Omega_0^2\, \mu\over 
c^2} = 3\times 10^{18}\, Z \left({\eta\over 0.1}\right) \left({\Omega_0\over 
10^4\; \hbox{s}^{-1}}\right)^2 \left({B_*\over 10^{12}\;\hbox{G}}\right) \hbox{eV},
\ee
where $\eta$ is the fraction of the voltage drop, $\Phi_{\rm wind}$, experienced 
by the charges, $B_*$ is the surface magnetic field strength (giving rise to
a magnetic moment $\mu\equiv B_*R_*^3$, in terms of the stellar radius $R_*$) and 
$\Omega_0$ is the initial pulsar rotation frequency.  In the case of PSR J1803-2137,
its measured spin-down age (Kassim and Weiler 1990) implies a magnetic field strength 
$B_*\approx 8.6\times 10^{12}$ G, and therefore $E_{max} \sim 24$ EeV.

As rotational energy is transferred to the particles, the pulsar spins down, and the 
injected particle energy drops.  As a result, the particles accelerated over the pulsar's 
spin-down lifetime produce a power-law distribution
\be 
N(E) = {9\over 4} {c^2 I \over Ze\mu} E^{-1}\;,
\ee
where $I$ is the neutron star's moment of inertia.  The fact that PSR J1803-2137 has a 
present day period of 133 ms, means that the distribution in this case extends down in 
energy only as far as $E_{min}\approx 6\times 10^{5}$ GeV, estimated from Equation (1) by 
setting $\Omega_0$ to its current value. 

The interaction of the accelerated, power-law (i.e., $N(E) = N_0 E^{-1}$) particles, 
assumed here to be protons, with an ambient medium of density $n_p$, leads to 
the production of pions and, subsequently, photons from the decay of neutral pions
and secondary lepton emission. Since Equation (2) directly yields the number of 
relativistic protons injected into the surrounding medium once the magnetic field
strength is specified, the cascade-induced emissivity depends solely on the properties
of the ambient medium, e.g., its number density. (A full description of the pp induced 
particle cascade and resulting broadband radiative emissivity is provided in Fatuzzo \& Melia 
[2003; hereafter FM03] and Crocker et al. [2005a], and will therefore not be reproduced 
in detail here.\footnote{It is worth pointing out, though, that FM03 use the empirical 
fits to the accelerator data based on a hybrid isobar/scaling model (Dermer 1986a,b).  
These fits are similar, but slightly different, from the cross sections used in 
other SNR treatments (see, e.g., Drury et al. 1994; Gaisser et al. 1998; Baring et al. 
1999).}) However, the rate at which these cosmic rays diffuse out of the scattering region 
depends on their rigidity and therefore on their energy. As a result, the energy distribution
of the particles remaining in the interaction zone will differ from the simple scaling
implied by Equation (2) and, moreover, this distribution evolves in time as the diffusion 
process differentiates between particles of different energy.

But one does not need to analyze the diffusion process in great detail to see that 
PSR J1803-2137 could not be the source of TeV photons in HESS J1804-216. 
The reason is that $E_{min}>> 1$ TeV, and regardless of how
the energy-dependent diffusion modifies the cosmic ray distribution at the source,
the photon spectrum below $E_{min}$ would have an index $\approx -1$, at odds with the
observed value $-2.7$.  One can see this in Figure~1, where we show the calculated photon 
spectrum in comparison with the TeV data, under the most favorable assumption that diffusion 
could somehow produce a proton distribution with index $-2.7$. The photon emissivity below 
$E_{min}$ is dominated by the $\pi^0$ cascade initiated by cosmic rays in the energy 
range $\sim 6\times 10^5-6\times 10^6$ GeV, and this is true for all proton power-law
indices $<0$. It therefore appears that under all circumstances, the contribution of 
PSR J1803-2137 to the TeV photon spectrum is too flat to account for the HESS data.

Even so, PSR J1803-2137 could still account for the putative cosmic ray anisotropy 
at $\sim$ EeV energies, without contributing measurably to the observed TeV flux, if 
the proton spectral index is $\sim -1$ (see Crocker et al. 2005a).  It should be noted,
however, that the production of neutrons within this scheme results from charge exchange
in $pp$ scattering.  A present day neutron production associated with particles
accelerated by PSR J1803-2137
would then require that the diffusion time for EeV protons (which would have been injected 
very early on by the pulsar) out of the surrounding region could not be much less than 
the pulsar age, and that the surrounding region have a density below $10$ cm$^{-3}$
so as to not violate the observed HESS data, as shown in Figure 1 (note that the 
photon emissivity scales directly with the ambient density).

\bigskip
\centerline{\sl 2.2 The SNR G8.7-0.1}
The expansion of an SNR into a dense molecular cloud
can produce shock accelerated protons whose interactions with that medium 
(via pp scattering) may also lead to an observable pionic decay signal above $\sim 100$ 
MeV (Drury et al.  1994; Sturner et al. 1997).  Indeed, this mechanism has been invoked 
to account for the association of several EGRET sources with SNRs interacting with their
dense surroundings (see, e.g., Gaisser et al. 1998; Baring et al. 1999; Berezhko \& V\"olk 
2000; Fatuzzo \& Melia 2005).  

In this section, we consider whether the broadband emission powered by SNR shocks 
in G8.7-0.1 can account for the HESS J1804-216 source without violating the
flux limits observed at other wavelengths.  We assume that shock acceleration 
within the SNR environment injects a power-law distribution of relativistic 
protons with a spectral index $\alpha = 2.0-2.4$ and maximum energy $E_{max}$ into 
a dense ($500$ cm$^{-3}$) shell at the SNR-cloud boundary (e.g., Chevalier 1999; 
Fatuzzo \& Melia 2005).  The value of $E_{max}$ can be approximated by taking the 
product of the remnant's age with the energy-gain rate for particles in a shock, 
\begin{equation}
\dot E(t)=10^8\,{B\,v_8^2(t)\over fR_J}\;\hbox{eV}\;{\hbox{s}}^{-1}\;,
\end{equation}
where $B$ ($\approx 10^{-5}$ G) is the magnetic field strength, $v_8$ is the
shock velocity in units of $10^8$ cm s$^{-1}$, $f\sim 10$ is the particle
mean free path along the magnetic field in units of its gyroradius, and
$R_J\sim 1$ is a factor that accounts for the orientation of the shock relative
to the magnetic field (Sturner et al. 1997). For the SNR age $T_{SNR}\sim 15,000$ years
(Kassim and Weiler 1990), and canonical values of the parameters, we estimate 
$E_{max} \sim 10^5$ GeV. Since this value exceeds the HESS date range for
HESS J1804-216, we ignore any high-energy truncation in the proton distribution 
when calculating the pion-decay spectrum.

In order to calculate the pp-initiated cascade leading to the gamma ray emissivity,
one must know the current relativistic proton distribution in the interaction region.
Once these particles leave the shock acceleration region, they diffuse into the
dense ($n\sim 500$ cm$^{-3}$) molecular environment where they scatter with the
low-energy ambient medium, though some eventually leave the system without scattering.
Given the $\sim 40$ mbarn pp scattering cross section, we estimate the cooling time
scale for these cosmic rays to be $\sim 70,000$ years, much longer than the age
of the remnant. (In reality, the pp scattering cross section increases slowly with
energy, such that this time scale shrinks to $\approx 15,000$ years for cosmic
rays with energy $\approx 1.7$ EeV. At this energy, however, the protons escape 
directly from the source.) As such, the injected proton distribution evolves 
primarily under the influence of energy-dependent diffusion, according to the 
simplified equation
\begin{equation}
{dN(E)\over dt}=Q(E)-{N(E)\over\tau(E)} \;,
\end{equation}
where $Q(E)$ is the injection rate and 
\begin{equation}
\tau\equiv {R^2\over D(E)}
\end{equation}
is the diffusion time scale, in terms of the system size $R$ and the diffusion 
coefficient $D(E)$. To simplify the procedure, yet retain the essential physics,
we solve this equation in two limits: (i) for energies such that $\tau(E)>>T_{SNR}$,
we put $N(E)=Q(E)\,T_{SNR}$; (ii) for energies such that $\tau(E)<<T_{SNR}$,
we take $N(E)=Q(E)\,\tau(E)$. The resulting power laws are connected smoothly at the
energy $E_{roll}$ where $T_{SNR}=\tau(E)$.

Unfortunately, the process of diffusion (here characterized by the diffusion
coefficient $D[E]$) is not well understood, and several viable prescriptions 
exist. To quantify the dependence of our results on the range of possibilities, 
we will use the formulation
\begin{equation}
D(E)\equiv \left({\lambda_{max}\,c\over 3\mu}\right)\left({E\over ZeB\,
\lambda_{max}}\right)^{2-\beta}\;,
\end{equation}
where $\lambda_{max}$ is the scale size of the largest magnetic fluctuations in the 
system (often similar to the scale size of the dynamical disturbance), $\mu$ ($\sim 1$)
represents the magnitude of the magnetic fluctuations relative to the underlying global
field, and $\beta$ is the index characterizing the fluctuation spectrum for the various
prescriptions of turbulence, i.e., $\beta=1$ for Bohm diffusion, $3/2$ for Kraichnan 
diffusion, and $5/3$ in the case of Kolmogorov diffusion.  In what follows, we shall 
examine the impact of all three forms of diffusion, and ascertain whether any may be 
ruled out for this system given the currently available data for HESS J1804-216. 

The actual value of $E_{roll}$ is rather sensitive to the choice of $R$ and
$\lambda_{max}$ (and to a lesser extent, $B$). For reasonable choices ($R\sim
1-10$ pc and $\lambda_{max}\sim 1-10 R$) of these variables, $E_{roll}$ ranges 
over several decades and spans the whole HESS energy range. As such,
$E_{roll}$ effectively functions as a free parameter, since neither $R$ nor
$\lambda_{max}$ are known precisely.

We show in Figures 2, 3, and 4 the gamma ray spectra of particle distributions
injected into the medium by shocks in SNR G8.7-0.1, and subsequently modified in energy
by diffusion. For all three cases of diffusion, the particle distribution index above 
$E_{roll}$ is $-2.7$, but in order to attain this value with the different $\beta$'s, 
the assumed proton injection power-law must be modified accordingly. A quick inspection 
of these three figures indicates that the principal impact of this feature is to alter 
the spectrum below $E_{roll}$, where the particles have not yet had time to diffuse 
through the medium. As such, the photon spectrum in this region is shaped by $Q(E)$. 
Thus, the spectrum below $E_{roll}$ is hardest for Bohm diffusion (Figure~2), and it 
softens progressively from Bohm to Kraichnan (Figure~3), and then to Kolmogorov (Figure~4).

Under the assumption of a uniform and steady injection rate $Q(E)$, as we have adopted
here, only the Bohm diffusion scenario results in a photon spectrum that does
not violate the EGRET upper limit. Because the ensuing emissivity scales as the product 
$N\cdot n_p$, the energy content of the relativistic particles is given by $U_E 
\approx 10^{49}\,(n_p/500\,{\hbox{cm}}^{-3})^{-1}$ ergs, which represents approximately 1\% 
of the supernova energy for the assumed density.  In this case (of Bohm diffusion),
the integrated flux between 100 MeV and 30 GeV is found to be $6\times 10^{-8}$ cm$^{-2}$ 
s$^{-1}$, and thus falls right at the EGRET threshold. In the near future, GLAST 
will greatly improve the sensitivity of measurements in this energy range, and will
therefore provide an important probe into the possible association of HESS J1804-216 with 
SNR G8.7-0.1.  In fact, the $\sim 1$ GeV emission should be detectable with GLAST after a 
one-year all-sky survey (indicated by the thick solid lines in Figures~2, 3, and 4)
regardless of which prescription of diffusion is active.  

For completeness we calculate the broadband emissivity for the case of Bohm diffusion
by including the flux contributions from secondary leptons 
(as outlined in detail in Fatuzzo \& Melia 2003). The decay of charged pions 
leads to the production of electrons and positrons in the scattering environment that 
in turn radiate via bremsstrahlung, synchrotron and inverse Compton scattering with the 
5.7 eV cm$^{-3}$ stellar photon field that permeates this region (Helfand et al. 2005; 
strictly speaking, this is the photon density at the galactic center, but given that we 
need it primarily to provide an upper limit, it will suffice as an estimate of the photon 
field at the location of G8.7-0.1 as well).  The resulting population of these secondary 
leptons depends on the injection rate (which is tied directly to the rate of pion decay, 
and subsequently, to the $\pi^0$ decay emissivity fixed by the HESS data) and the energy 
loss rates from the aforementioned emission processes and from Coulomb processes.  The 
cooling time $E/\dot E$ for each of these four processes is shown in Figure~5 for an 
assumed ambient density $n_p = 500$ cm$^{-3}$ and magnetic field strength $B=10^{-5}$ G.  
It is clear that leptons with energy between $\sim 10^2$ and $10^7$ MeV will not have 
time to cool during the remnant's estimated lifetime.  The lepton distribution is 
therefore approximated by using the lower part of the steady-state lepton distribution 
curve and the secondary lepton injection rate curve multiplied by the age of the remnant.  

The broadband spectrum (from radio to TeV energies) for the case of Bohm diffusion is 
shown in Figure~6. This now includes the various contributions from the secondary leptons 
as well as photons emitted during pion decays. Clearly, the radiative flux produced by 
secondary particles falls well below all broadband limits.  It is worth noting, however, 
that a secondary lepton population created by the proton distribution with spectral index
of $\sim -2.3$ (also mirrored by the injected secondaries) can reproduce the observed 
radio emission if the magnetic field strength is $\sim 0.3$ mG.  Such field strengths 
have been associated with SNR - molecular cloud interactions (Claussen et al. 1997; 
Koralesky et al. 1998; Brogan et al. 2000).  However, this choice of spectral index 
for the protons is clearly at odds with the EGRET observations.

\section{Conclusion}
We are left with a rather intriguing situation. On the one hand, we have shown that the TeV 
emission associated with the new source HESS J1804-216 may be a signature of particle 
acceleration and injection into the ambient medium by the shell of SNR G8.7-0.1, but
not by the pulsar PSR J1803-2137 embedded within it. Reasonable values of the physical 
variables are sufficient to account for its radiative characteristics, and associated 
flux limits observed from this region at other wavelengths.

On the other hand, the maximum energy of cosmic rays energized by the presumed shock
in G8.7-0.1 is severely limited by the acceleration efficiency and time, and would
seem to be restricted to values below about $10^5$ GeV. This scenario would not support
the viability of a CR anisotropy from this region, and would be consistent with the
findings of the Auger Observatory, which does not confirm earlier claims based on
the SUGAR and AGASA data. 

The pulsar model for HESS J1804-216 does not work at TeV energies, but unlike the
SNR source, the energy of particles accelerated near the pulsar's light cylinder 
reach values as high as $\sim 24$ EeV. The capability of PSR J1803-2137 to produce 
such energetic cosmic rays means that a CR anisotropy in the direction of HESS J1804-216 
could be accounted for with this model.  Such a scenario would require that EeV protons
produced nearly 15,000 years ago have not all diffused out of the surrounding, 
lower density ($< 10$ cm$^{-3}$) region.  In contrast, particles associated with 
the HESS source would result from the SNR-cloud interaction, and populate a 
denser (and smaller) shell.  Such an anisotropy, though, would require the 
unlikely circumstance of Auger's findings being incorrect. 

To see how these considerations impact the broader context of high-energy
activity at the galactic center, let us firstly take it that, as the AUGER data 
suggest, there really is (currently) no EeV CR anisotropy in this direction. 
Then we are led to one of the two following conclusions: 
\begin{enumerate}
\item if the SUGAR and AGASA data and the analyses thereof (suggesting the existence 
of galactic center CR anisotropies) are broadly correct, then a source both variable 
on decadal timescales and capable of accelerating particles to beyond $10^{18}$ eV 
exists. Only a source associated with a compact object, most likely a pulsar, could 
satisfy these requirements but, as may be seen above, it is then difficult to conceive 
of a source that might produce both the required variability and a power-law spectrum 
of accelerated particles; it would seem, then, that the HESS J1804-216 source and the 
SUGAR anisotropy cannot be directly attributed to the same object. Likewise, the 
high-energy galactic center source scenarios described in Crocker et al. (2005a) could 
not account for the required variability plus power-law spectrum of accelerated particles.
\item if, on the other hand, the SUGAR and AGASA data and/or the analyses thereof are in error, 
then the scenarios outlined in Crocker et al. (2005a)---which are predicated on the 
SUGAR and AGASA anisotropies being real---cannot hold. There also necessarily exists
an implicit upper limit on the strength of any $10^{18}$ eV neutron source located 
within the field of view of AUGER and, in particular, towards the galactic center. 
Furthermore, the in-situ high-energy hadronic population inferred on the basis of the 
TeV radiation detected by the HESS instrument from the galactic center (and also from 
the J1804-216 source) must cut off below $\sim$ EeV. Alternatively, it may be that,
even if this population does continue up to the EeV energy scale as an undistorted 
power law, the simple scaling behavior for the $pp \to n X$ type interaction assumed by
Crocker et al. (2005a) and also implicitly employed above (given the dearth of direct 
experimental data on neutron production at these very high center-of-mass energies) 
fails and, in fact, significantly over-estimates neutron production (Grasso and Maccione 
2005).

\end{enumerate}

Should the analysis of the currently available AUGER data be incomplete, so that, in 
particular, the galactic center (angular) region be CR over-abundant at $\sim$ EeV 
energies with an amplitude suggested by the AGASA and SUGAR data and consequent 
analyses, a proton acceleration model with PSR J1803-2137 as the source is a viable
explanation for the SUGAR CR anisotropy. We note in this regard that the minimum energy
associated with this pulsar's particle injection is well beyond the HESS data range, so 
this eventuality would not be impacted by the low energy data. Alternatively, 
it may be that, as discussed by
Crocker et al. (2005a), the SUGAR point source be real but actually located at the 
galactic center and, therefore, associated not with HESS J1804-216, but rather the 
galactic center source detected by HESS (this would, of course, require that SUGAR's 
directional determination be in error by $\sim 8^\circ$, but then the problem with 
AGASA's non-observation of the SUGAR point source would be resolved).

Pulsars such as PSR J1803-2137, embedded within the environment surrounding the
SNR G8.7-0.1, can be viable sources of EeV cosmic rays within the Galaxy. This
warrants further analysis. In future work, we will examine whether the known
population of sources such as this, distributed throughout the Milky Way, can
conceivably account for the isotropic component of the CR spectrum observed at
$\sim$ EeV energies.

\bigskip 
\centerline{\bf Acknowledgments} 

This research was partially supported by NSF grant AST-0402502 (at Arizona). 
FM is very grateful to the University of Melbourne for its support (through a 
Miegunyah Fellowship).  MF is supported by the Hauck Foundation through Xavier 
University.

\newpage
\begin{figure}
\figurenum{1}
{\epsscale{0.9} \plotone{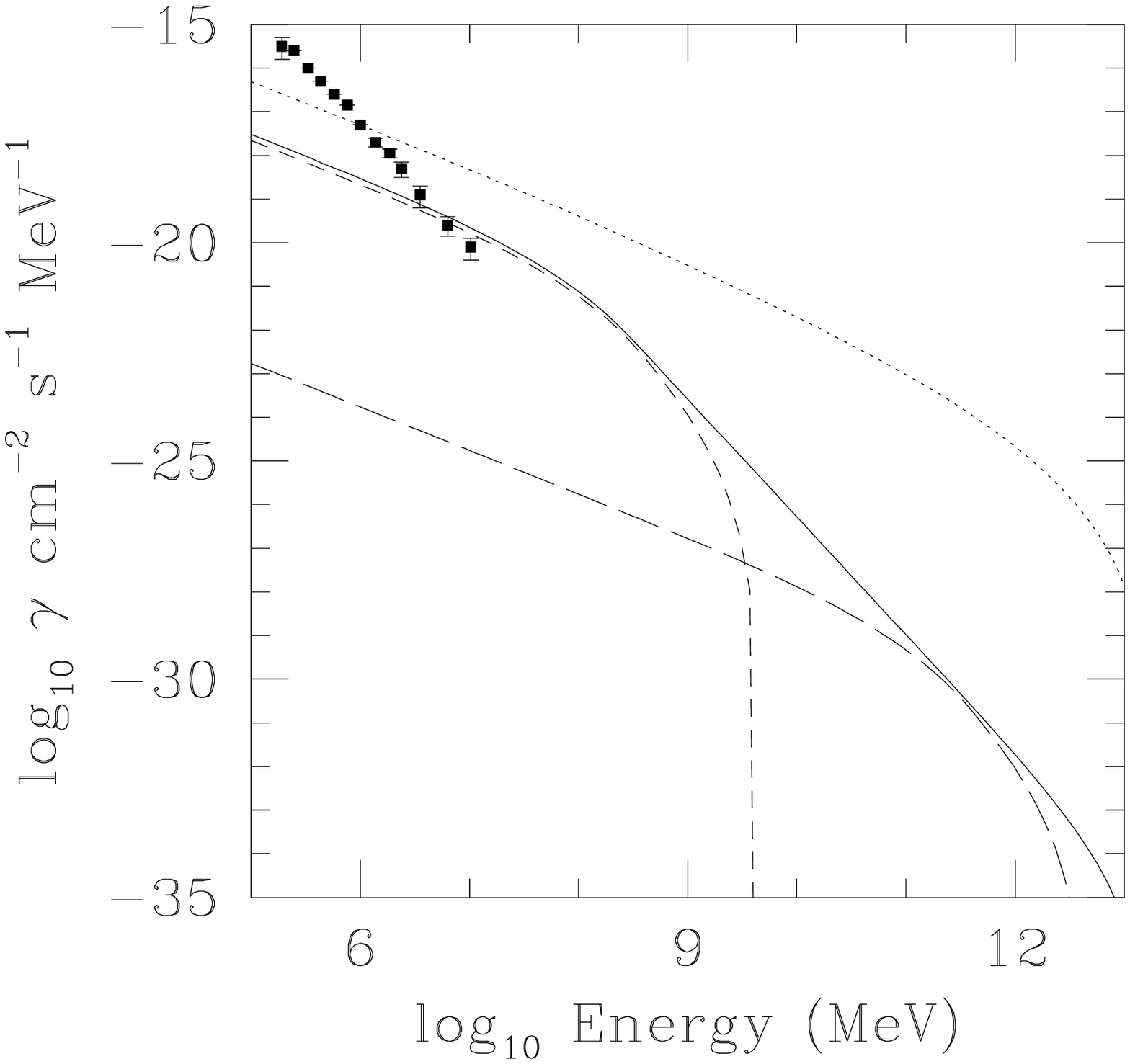} }
\figcaption{Illustrative $\gamma$-ray emissivity from the pulsar-powered pion-decay model
with $E_{max}\sim 24$ EeV and $E_{min}\sim 6\times 10^5$ GeV (both calculated from
its inferred spin-down age) and an assumed ambient density of $10^3$ cm$^{-3}$.
The dotted curve shows the resulting emissivity if the injected particles do not
diffuse out of the region.  The solid curve shows the resulting emissivity
for an idealized distribution that, as a result of diffusion, has a spectral 
index of -2.7. In both cases, the $\pi^0$ cascade induced by protons with energy $E$
produces a photon spectrum with index $\approx -1$ below $E_{min}$.  In the case where
diffusion acts to remove high-energy particles from the emission region, 
the spectrum below $E_{min}$ is almost entirely due to protons with 
energy $\sim E_{min}$ (whose contributions are shown by the short-dashed curve). 
Similarly, the long-dashed curve shows the photon spectrum
resulting from the $\pi^0$ cascade initiated by protons at $E\approx 6\times 10^{8}$ GeV.
Regardless of how much the energy-dependent diffusion modifies the injected cosmic ray
spectrum above $E_{min}$, the pulsar model therefore cannot account for the steep
spectrum (index $-2.7$) measured by HESS.}
\end{figure}

\newpage
\begin{figure}
\figurenum{2}
{\epsscale{0.9} \plotone{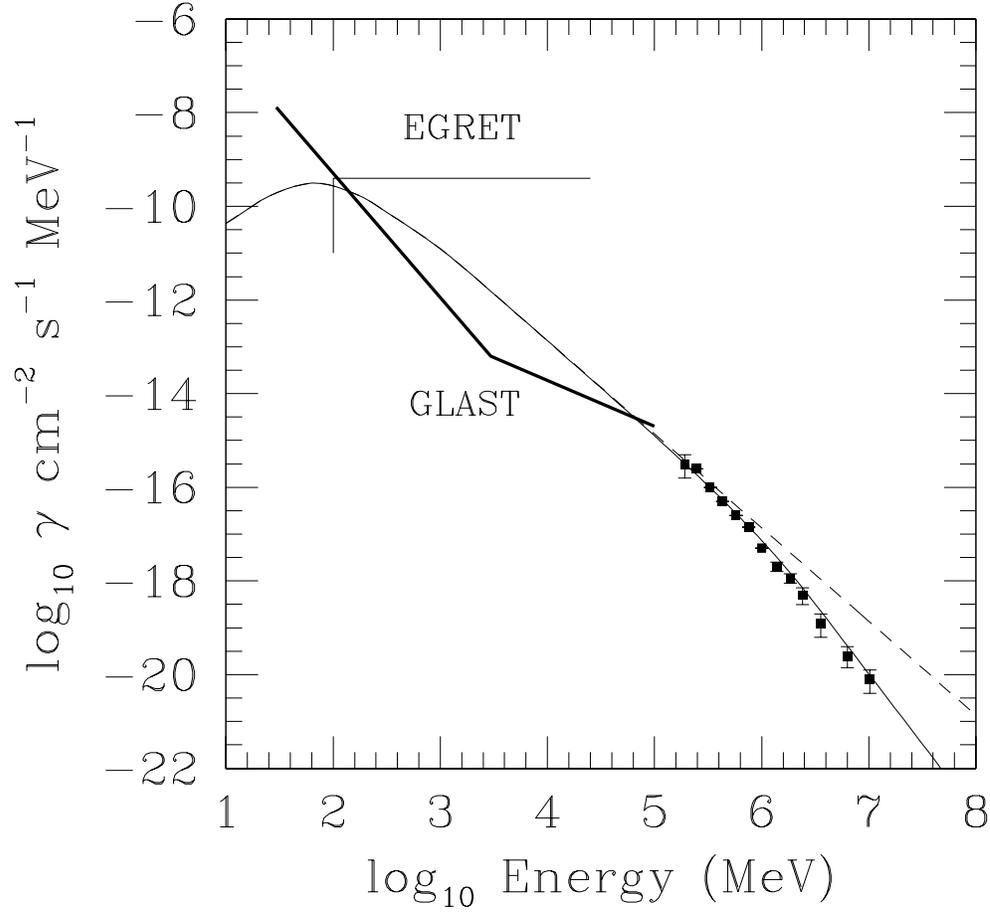} }
\figcaption{The $\gamma$-ray emissivity (solid line) for a particle distribution 
injected with index $-2.0$ and modified by Bohm diffusion (and $E_{roll}=7\times 10^3$
GeV), and the corresponding spectrum (dashed curve) produced without diffusion. 
The EGRET bar is an upper limit, and the GLAST curve is the simulated one-year 
all sky survey limit.}
\end{figure}

\newpage
\begin{figure}
\figurenum{3}
{\epsscale{0.9} \plotone{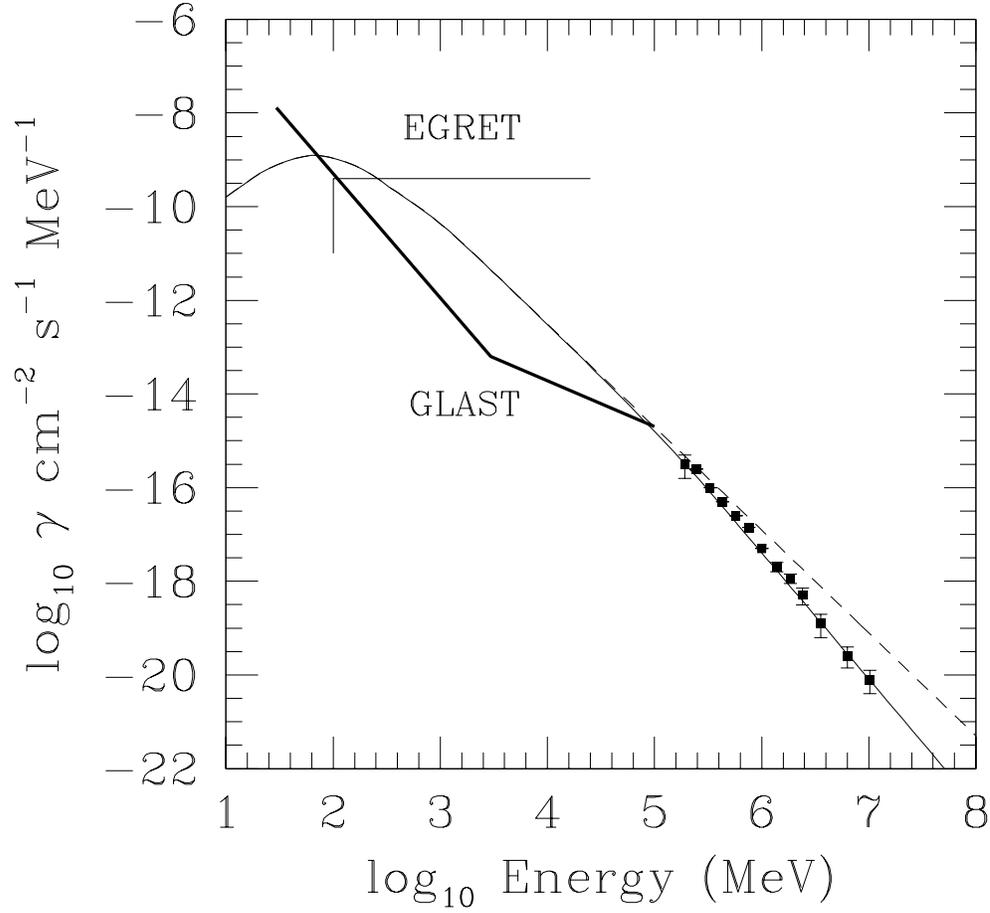} }
\figcaption{Same as Figure~2, except now for Kraichnan diffusion with $E_{roll}=
10^3$ GeV, and an injected particle distribution index $-2.2$.}
\end{figure}

\newpage
\begin{figure}
\figurenum{4}
{\epsscale{0.9} \plotone{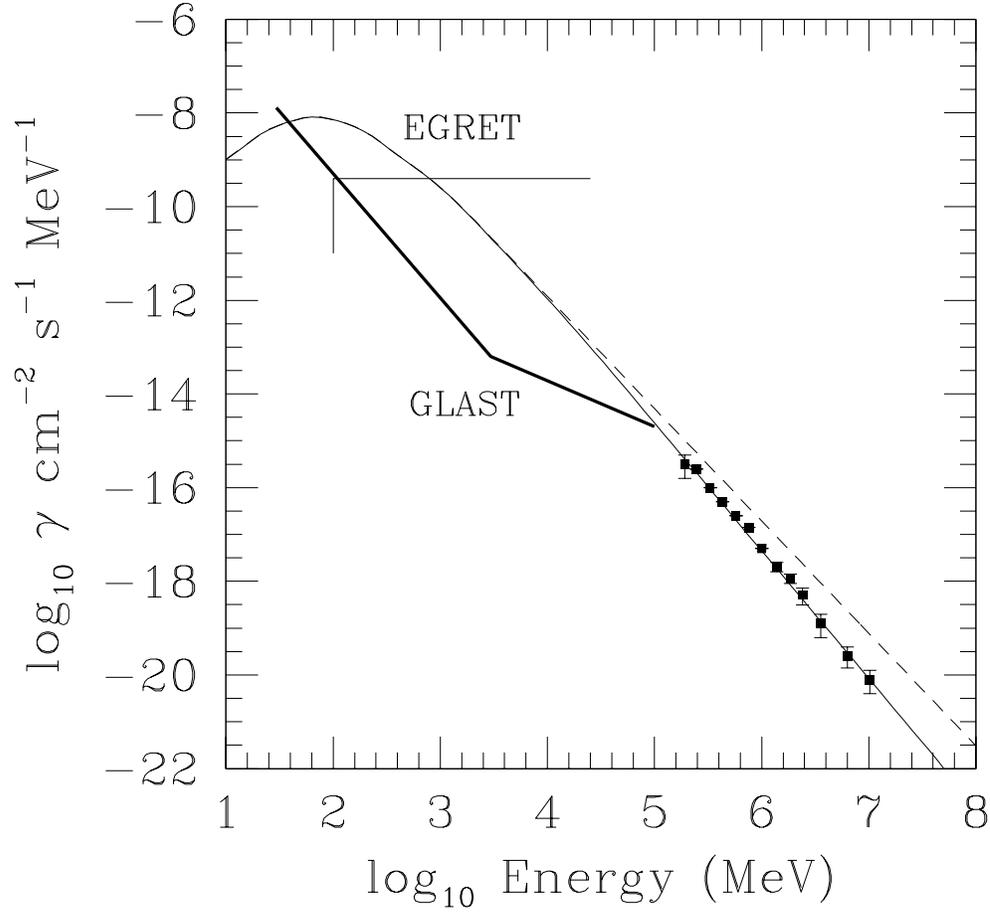} }
\figcaption{Same as Figures~2 and 3, except now for Kolmogorov diffusion
with $E_{roll}=100$ GeV, and an injected particle distribution index $-2.4$.}
\end{figure}

\newpage
\begin{figure}
\figurenum{5}
{\epsscale{0.9} \plotone{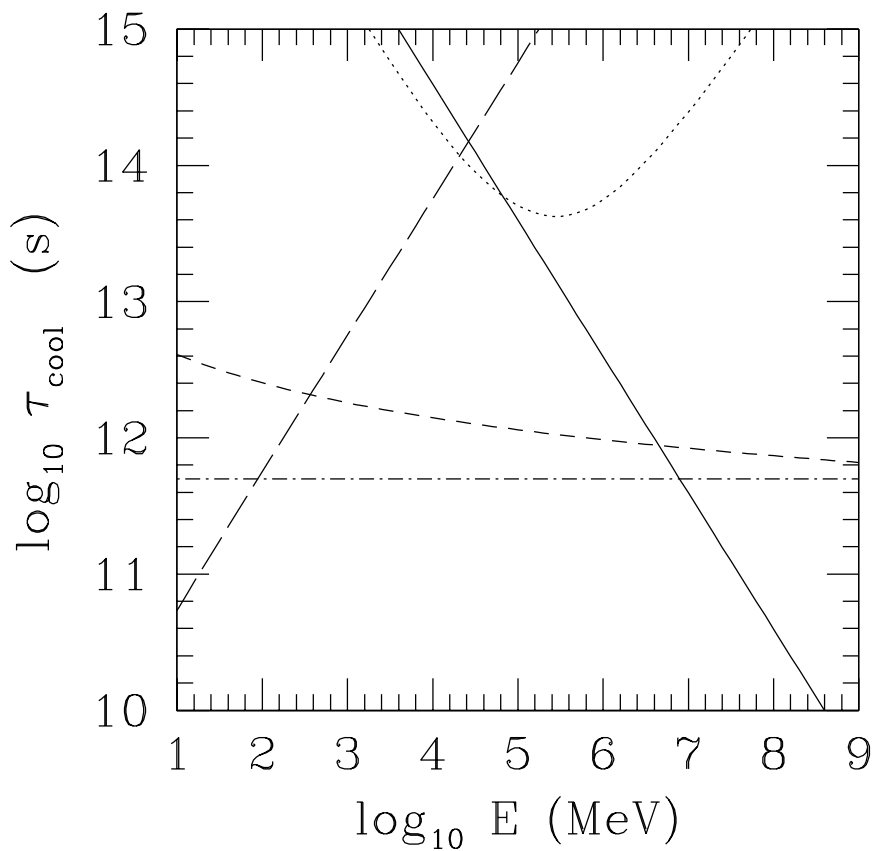} } 
\figcaption{The cooling time $\tau = E / \dot E$ as a function of energy 
for leptons interacting with a medium of density $n_H = 500$ cm$^{-3}$ 
and magnetic field strength $B = 10^{-5}$ G.  Short dashed curve: bremsstrahlung 
cooling; solid curve: synchrotron cooling; long dashed curve: Coulomb losses; 
dotted curve: inverse Compton scattering with the ambient stellar photon field.}
The dot-dashed curve represents the value of the SNR age.
\end{figure}

\newpage
\begin{figure}
\figurenum{6}
{\epsscale{0.9} \plotone{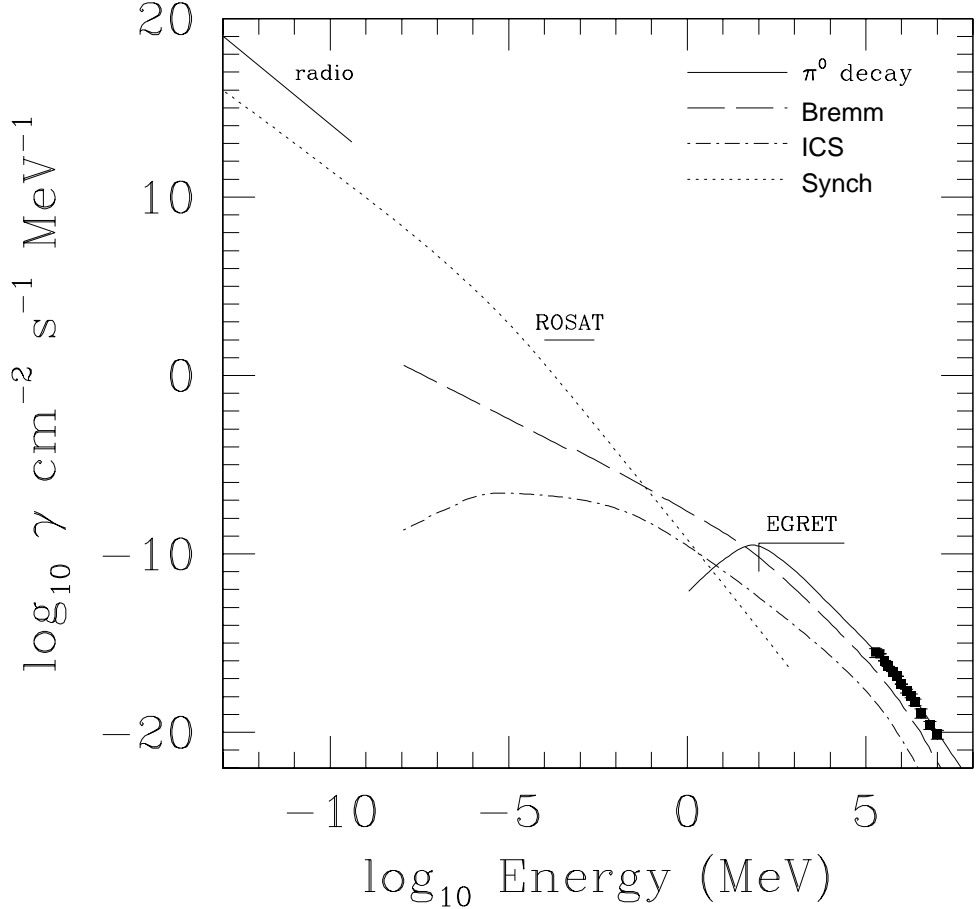} } 
\figcaption{The $\gamma$-ray emissivity from the SNR-powered pion-decay 
model with $n_H = 500$ cm$^{-3}$, and $B = 10^{-5}$ G.  
The remnant's age is assumed to be $15,000$ years in order to determine the secondary
leptons' (non steady-state) distribution. Solid line: photons produced via the decay 
of neutral pions; long dashed curve: bremsstrahlung emission from the secondary 
leptons; dot-dashed line: inverse Compton scattering emission from the secondary 
leptons interacting with the background stellar field; dotted line:  synchrotron 
emission from the secondary leptons.  The HESS data are represented as dark squares.
Also shown are the EGRET upper limit, and the measurements made with ROSAT and the
VLA, both of which must be considered as upper limits as well, given the differences
in field-of-view.}
\end{figure}

\end{document}